\documentclass[12pt]{article}
\usepackage{amsmath,amssymb}
\usepackage{graphicx}
\usepackage{caption}
\usepackage{color}
\usepackage{dcolumn}
\usepackage{bm} 
\usepackage[numbers,super,comma,sort&compress]{natbib}
\usepackage[T1]{fontenc}
\definecolor{background-color}{gray}{0.98}
\newcommand{\ket}[1]{|#1\rangle}
\newcommand{\bra}[1]{\langle#1|}
\newcommand{\Tr}{\text{Tr}}
\newcommand{\Liouvillian}{\mathcal{L}}

\title{Quantum nonlocality in the excitation energy transfer in the Fenna-Matthews-Olson complex}
\author{Charlotta Bengtson\thanks{Department of Chemistry - {\AA}ngstr\"om Laboratory, 
Theoretical Chemistry, Uppsala University, Box 538, SE-751 21 Uppsala, Sweden}, \ \ 
Michael Stenrup\thanks{Department of Chemistry - {\AA}ngstr\"om Laboratory, 
Theoretical Chemistry, Uppsala University, Box 538, SE-751 21 Uppsala, Sweden} 
\thanks{Uppsala Center for Computational Chemistry - UC$_3$, Uppsala University, 
Box 538, SE-751 21 Uppsala, Sweden}, \ \ 
Erik Sj\"oqvist\thanks{Department of Physics and Astronomy, Uppsala University, Box 516, 
SE-751 20 Uppsala, Sweden}}
\begin{document}
\maketitle
\begin{abstract}
The Fenna-Matthews-Olson (FMO) complex - a pigment protein complex involved in photosynthesis 
in green sulfur bacteria - is remarkably efficient in transferring excitation energy from light harvesting 
antenna molecules to a reaction center. Recent experimental and theoretical studies suggest that 
quantum coherence and entanglement may play a role in this excitation energy transfer (EET). We 
examine whether bipartite quantum nonlocality, a property that expresses a stronger-than-entanglement 
form of  correlation, exists between different pairs of chromophores in the FMO complex when 
modeling the EET by the hierarchically coupled equations of motion method. We 
compare the results for nonlocality with the amount of bipartite entanglement in the system. In 
particular, we analyze in what way these correlation properties are affected by different initial 
conditions. It is found that bipartite nonlocality only exists when the initial conditions are chosen in 
an unphysiological manner and probably is absent when considering the EET in the FMO 
complex in its natural habitat. It is also seen that nonlocality and entanglement behave quite 
differently in this  system. In particular, for localized initial states, nonlocality only exists on a 
very short time scale and then drops to zero in an abrupt manner. As already known from previous 
studies, quantum entanglement between chromophore pairs on the other hand is oscillating and 
exponentially decaying and follow thereby a pattern more similar to the chromophore population 
dynamics. The abrupt disappearance of nonlocality in the presence of nonvanishing entanglement 
is a phenomenon we call {\it nonlocality sudden death}; a striking manifestation of the 
difference between these two types of correlations in quantum systems. 
\end{abstract}
\clearpage
\makeatletter
\renewcommand\@biblabel[1]{#1.}
\makeatother
\bibliographystyle{apsrev}
\renewcommand{\baselinestretch}{1.5}
\normalsize
\clearpage
\section*{\sffamily \Large INTRODUCTION} 
In recent years, there has been an increasing interest in studying biological systems in
terms of the existence of nontrivial quantum effects 
\cite{Adolphs2006,Engel2007,Ishizaki2009a,Wilde2010,Sarovar2010,Hoyer2010,Caruso2010,Hoyer2012}.
Especially the excitation energy transfer (EET) between chromophores in photosynthetic complexes 
has been heavily studied ever since quantum coherence between excited states in the 
Fenna-Matthews-Olson (FMO) complex was experimentally verified by two dimensional 
electronic spectroscopy \cite{Engel2007}. Not only was the existence of coherent pathways 
in the FMO complex revealed, quantum coherence was also shown to last much longer than 
expected for such large and noisy systems, and a hypothesis is that this may play a role in the 
known very efficient EET in photosynthetic complexes \cite{Sarovar2010, Plenio2013}. Since then,
several attempts to explain why photosynthetic complexes could benefit from coherent EET 
have been made \cite{Caruso2010,Hoyer2012}. One such is that the quantum coherence
could help to direct the energy flow towards the reaction center in a unidirectional manner
\cite{Ishizaki2009a,Hoyer2012}. 

The experimental evidence for coherent EET in the FMO complex motivates the development 
of new methods to model the excited state dynamics of chromophores in a surrounding
protein scaffold. It has become clear that Markovian models cannot fully capture the behaviour
of the system-environment interactions in these systems and hence non-Markovian models 
have to be considered. Based on the hierarchial expansion technique proposed in
Refs.~\cite{Tanimura1989,Tanimura1990}, Ishizaki and Fleming refined the theoretical
framework and developed a tool for investigating the excited state dynamics in a photosynthetic
complex \cite{Ishizaki2009a,Ishizaki2009b}. This form of hierarchially coupled equations of 
motions (HEOM) has served as the benchmarking method for these systems since it is able 
to describe quantum coherent motion and incoherent electron hopping in a unified manner 
\cite{Ishizaki2009b}. Using this method, Sarovar {\it et al.} \cite{Sarovar2010} examined 
quantum entanglement in the FMO complex and found that bipartite entanglement of 
chromophores exists on a time scale relevant for the mechanism of EET. Since entanglement 
can be used as a resource in quantum information processing tasks, it was speculated if the 
findings could play a role in explaining the near unity efficiency for these complexes to 
convert solar light into chemical energy.

Quantum nonlocality, as introduced by Bell \cite{Bell1964}, is another correlation property 
of composite quantum systems. While quantum entanglement, already mentioned by 
Schr\"{o}dinger as the key property that distinguishes quantum mechanics from the classical 
world \cite{schrodinger1935}, is a meaningful property only for quantum systems, nonlocality 
is given by a criterion that is formed without specifying whether the systems are quantum 
mechanical or not. In the quantum-mechanical context, although entanglement and nonlocality 
are the same for pure states, this is not in general true for mixed states, i.e., they are inequivalent 
properties for composite quantum systems \cite{Horst2013,Augusiak2015}. In particular, 
nonlocality is a stronger correlation property than entanglement in the sense that nonlocality 
implies entanglement but not vice versa \cite{Werner1989}.

The aim of this study is to examine whether bipartite nonlocality can exist during EET in 
the FMO complex, which would imply a striking nonclassical feature of this process. By 
studying the existence of quantum nonlocality in such a complex, it is possible to add 
insights into if and how quantum effects play a role in photosynthesis; insights that 
could be of interest for artificial photosynthesis and solar cells as well as for quantum 
computation. In recent studies \cite{Zhang2015,Creatore2013}, a considerable enhancement 
in the efficiency due to quantum coherence was found in model systems of photosynthesis 
mimicking solar cells. If such discoveries also can be connected to the existence of entanglement 
and nonlocality, it would be desirable to explore the underlying physics of such phenomena 
further.

\section*{\sffamily \Large MODELING THE EET IN THE FMO COMPLEX} 
\label{sec:model}
EET in the FMO complex has been modeled by employing the HEOM method \cite{Ishizaki2009a,Ishizaki2009b}. A description of the FMO complex as well as the quantities 
and conditions used to model EET, is given followed by a brief review of the HEOM based 
model of the FMO complex, modified by including an explicit mechanism of the excitation energy 
trapping at the reaction center.

\begin{figure}[ht]
\centering
\includegraphics[width=0.6\textwidth]{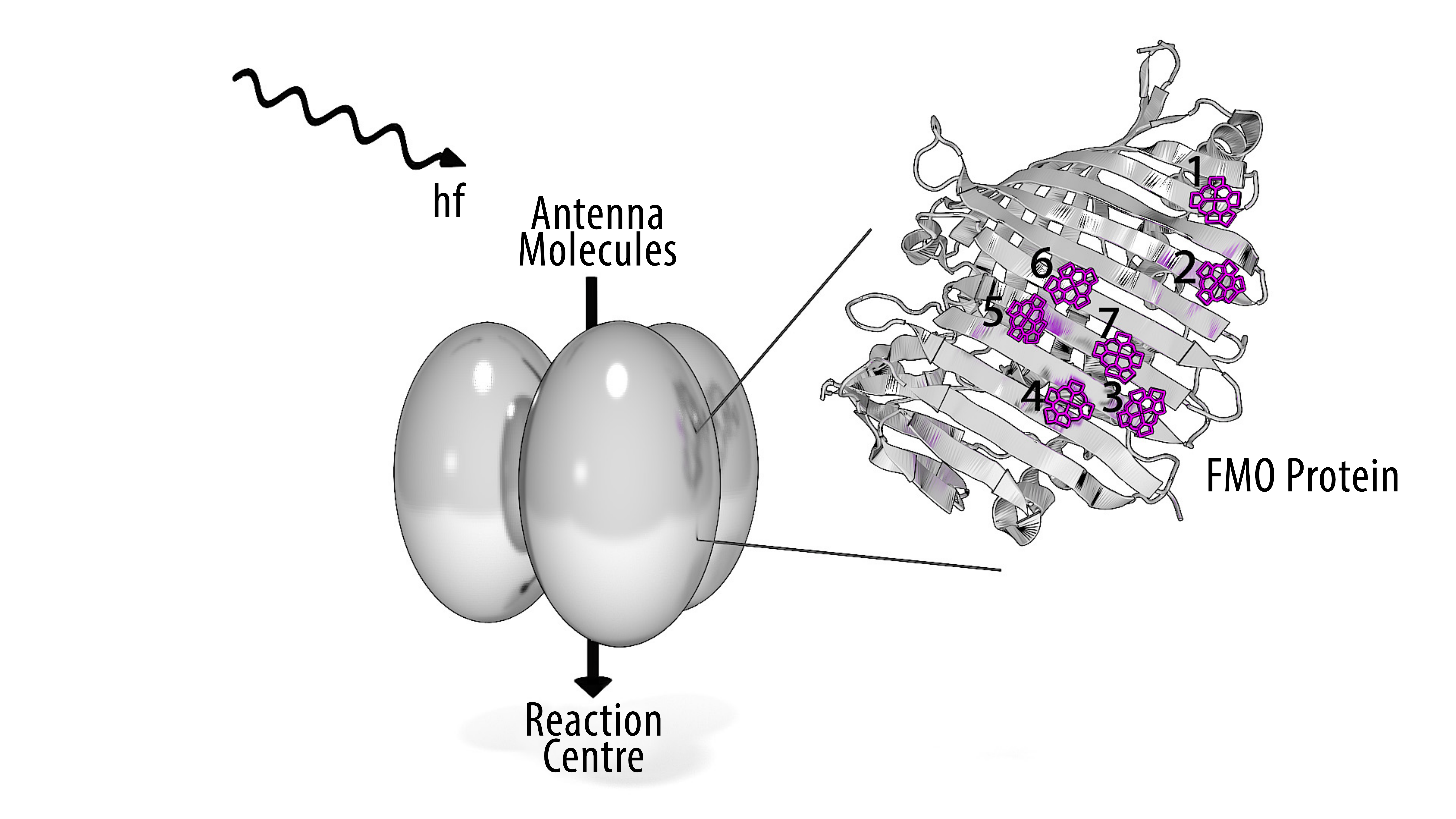}
\caption{The FMO complex trimer and its position between the antenna molecules and the reaction 
center to the left, and the chromophores in their protein scaffold for one monomer to the right. 
The chromophores are chlorophylls in this particular system.}
\label{FMO}
\end{figure}

\section*{\sffamily \Large The FMO complex} 
\label{sec:fmo}
The FMO complex is a photosynthetic complex found in green sulfur bacteria, {\it Chlorobaculum
tepidum}. It consists of three identical monomers, each containing seven chromophores embedded 
in a protein scaffold, as can be seen in Fig. \ref{FMO}. The structure of this complex as well as its site 
excitation energies and inter-site couplings have been experimentally investigated by different 
spectroscopic methods \cite{Engel2007,Panitchayangkoon2010,Wen2009} and the system has been 
used as a model for larger light harvesting complexes for more than 35 years \cite{Pearlstein1978}.

The FMO complex does not include any antenna molecules. In other words, it is not responsible 
for capturing the light energy, but only for transferring it to the reaction center. The antenna 
molecules are located in such a way that they can transfer excitation energy to the three monomer 
units. We restrict the system studied to one monomer; a reasonably simplifying assumption in 
the description of correlations accompanying EET in this system.

The individual chromophores in the FMO complex can be modeled as two level systems (TLSs) by 
taking into account only the $S_0 \rightarrow S_1$ transition, with $S_0$ and $S_1$ being 
the singlet ground state and singlet first excited state, respectively. Furthermore, since 
Green sulfur bacteria recieve very little sunlight in their natural habitat \cite{Sarovar2010}, it is 
reasonable to assume that the FMO complex only contains one such excitation at a time. This 
reduces the electronic Hamiltonian of the FMO complex (one monomer) to have the form
\begin{eqnarray}
\label{hamiltonian}
\hat{H}_e = \sum_{k=1}^7 E_k \ket{k}\bra{k} + \sum_{k\neq l}^7 J_{kl}\ket{k}\bra{l} ,
\end{eqnarray}
where $E_k$ is the electronic excitation energy of chromophore $k$ when being in its excited
state, while the other chromophores remain in their ground state. This corresponds to 
the localized state   
\begin{eqnarray} 
\ket{k}=\ket{S_{0}^{1} \ldots S_{1}^{k} \ldots S_{0}^{7}} , 
\label{eq:sitebasis}
\end{eqnarray}
where the superscript denotes the chromophore site. Furthermore, $ J_{kl}$ is the dipole coupling  
describing the electrostatic interaction between the charge distribution of chromophores 
$k$ and $l$. It depends strongly on the dipole moment orientations and relative positions 
of the chromophores in the protein complex structure \cite{Adolphs2006}. 

Calculations suggest that the most favourable structure is that chromophores 3 and 4 are 
linked directly to the reaction center \cite{Adolphs2006}. This result has been confirmed 
experimentally when examining the FMO complex with masspectrometry \cite{Wen2009}. 
Since the structure of the FMO is known \cite{Adolphs2006}, this implies that chromophores 
1 and 6 connect to the base plate (that is the part of the antenna molecule complex that 
connects it to the FMO). The simplest way to model EET in the FMO complex would hence 
be to assume a localized excitation on either chromophore 1 or 6 as the initial conditions for 
EET. This can be written as
\begin{eqnarray}
\hat{\varrho}_{\textrm{localized}}^{x} = \ket{x} \bra{x}, \ \ x=1,6.
\label{ic_localized}
\end{eqnarray}
However, since the distance between the antenna molecules and the base plate as well 
as the distance between the base plate and the FMO are so large  ($\sim$2 nm  
\cite{Huh2014,Martiskainen2012,Pedersen2010} and $\sim$1.5 nm \cite{Huh2014}, 
respectively) in comparison to the intracomplex distances, a model that better captures  
the condition for initial excitation of the FMO complex in its natural habitat is to assume 
that the excitation is transferred from the base plate to the FMO 
by F\"{o}rster resonance energy transfer (FRET) \cite{Leon-Montiel2014}. This would populate the 
FMO exciton states $\ket{e_{r}}$, being the eigenstates of the Hamiltonian in Eq.~(\ref{hamiltonian}), 
in proportion to their occurrence at chromophore 1 or 6. These FRET initial states can hence be 
written as
\begin{eqnarray}
\hat{\varrho}_{\textrm{FRET}}^{x} = 
\sum_{r=1}^7 \left| \langle{x} \ket{e_{r}} \right|^2 \ket{e_{r}} \bra{e_{r}}, \ \ x=1,6. 
\label{ic_fret}
\end{eqnarray}
In this work, we shall examine correlations between chromophore pairs in EET arising from 
the initial states given by Eqs.~(\ref{ic_localized}) and (\ref{ic_fret}). 

\section*{\sffamily \Large The HEOM method for the FMO complex} 
\label{sec:heom}
HEOM \cite{Ishizaki2009a,Ishizaki2009b} is a numerically exact method in which the environmental 
influence on a quantum system is treated in a statistical manner. For the FMO complex, the system 
of interest is the seven chromophores in one of the monomers with the Hamiltonian given in 
Eq.~(\ref{hamiltonian}). The protein scaffold is modeled as a set of harmonic oscillator modes, 
i.e., a phonon bath, which represents nuclear motion, both intramolecular and those of the 
environment. The transfer of excitation energy from one chromophore to another occurs via 
nonequilibrium nuclear configuration (phonon states) in accordance with a vertical 
Franck-Condon transition. The phonons, locally and linearly coupled to each chromophore 
$k$, relax to their equilibrium states under energy loss. This so-called reorganization energy, 
denoted $\lambda_k$, can be measured via the Stokes shift \cite{Ishizaki2009a,Ishizaki2010}.

The memory of the local environment of chromophore $k$ is characterized by a relaxation
function $\Gamma_k (t)$, modeled as an overdamped Brownian oscillator, which takes the form
\begin{eqnarray}
\Gamma_k (t) = 2 \lambda_k e^{-\gamma_k t}.
\end{eqnarray}
The parameter $\gamma_k$ represents the time scale of the fluctuations and energy dissipation 
for the $k$th chromophore and is related to the non-Markovian behaviour of the dynamics. The 
relaxation function, and hence $\gamma_k$, can be investigated by two dimensional electronic 
spectroscopy \cite{Ishizaki2009a,Ishizaki2010}.

The time dependent density operator $\hat{\varrho}(t)$ describing EET in the FMO complex is 
obtained by solving a set of hierchically coupled equations of motion,
\begin{eqnarray}
\label{heom}
\frac{\partial}{\partial t} \hat{\zeta}(\textbf{n},t) =
- \left( i \Liouvillian_e + \sum_{k=1}^7 n_k \gamma_k \right)
\hat{\zeta}(\textbf{n},t) + 
\sum\limits_{k=1}^7 \left[ \Phi_k \hat{\zeta}(\textbf{n}_{k+},t) + n_k \Theta_k
\hat{\zeta}(\textbf{n}_{k-},t) \right],
\end{eqnarray}
where the operator $\hat{\zeta}(\textbf{0},t)$ is identical to $\hat{\varrho}(t)$ while the 
higher order operators $\hat{\zeta}(\textbf{n} \neq \textbf{0},t)$ are auxiliary operators. Here, 
\textbf{n} is a set of nonnegative integers, $\textbf{n}=(n_1,n_2,..,n_7)$. The notation 
$\textbf{n}_{k+} (\textbf{n}_{k-})$ stands for adding (substracting) 1 to the corresponding 
$n_k$ in \textbf{n}. The Liouvillian superoperator $\Liouvillian_e$ is composed of the electronic
Hamiltonian of the FMO complex and the site reorganization energies, and takes the form
\begin{eqnarray}
\Liouvillian_e \hat{g} = [ \hat{H}_e + \sum_{k=1}^7 \lambda_k \ket{k}\bra{k} , \hat{g} ]
\end{eqnarray}
with $[\cdot,\cdot]$ the commutator and $\hat{g}$ any linear operator acting nontrivially on 
the single excitation subspace of the full Hilbert space of the FMO complex. The superoperators 
$\Phi_k$ and $\Theta_k$ are so called phonon-induced relaxation operators and correspond 
to the influence of the environmental fluctuations. They have the form
\begin{eqnarray}
\Phi_k \hat{g} & = & i [\hat{V}_k,\hat{g}]
\end{eqnarray}
and
\begin{eqnarray}
\Theta_k \hat{g}  & = & i\frac{2 \lambda_k}{\beta \hbar^2} [\hat{V}_k, \hat{g}] +
\frac{\lambda_k}{\hbar} \gamma_k \{ \hat{V}_k, \hat{g} \}
\end{eqnarray}
with $\hat{V}_k = \ket{k} \bra{k}$ and $\{ \cdot ,\cdot \}$ the anti-commutator. 
Here, $\beta=(k_BT)^{-1}$ with $k_B$ being the Boltzmann constant and $T$ the temperature 
of the phonon bath.

The HEOM can be truncated by setting
\begin{eqnarray}
\frac{\partial}{\partial t} \hat{\zeta}(\textbf{n},t) = -i \Liouvillian_e\hat{\zeta}(\textbf{n},t) ,
\end{eqnarray}
for all \textbf{n} satisfying $n_1+n_2+ \ldots +n_7=N$, $N$ being the truncation level. This 
condition will terminate the generation of auxiliary operators.

Following Ref. \cite{Shabani2012}, a Liouvillian that models the trapping of excitation of 
chromophores 3 and 4 is added to the right-hand side of Eq.~(\ref{heom}).
It has the form
\begin{eqnarray}
\Liouvillian_\mathrm{trap} \hat{\zeta}(\textbf{n},t) =
-r_\mathrm{trap} \{\ket{3} \bra{3},\hat{\zeta}(\textbf{n},t) \}
-r_\mathrm{trap}\{\ket{4} \bra{4},\hat{\zeta}(\textbf{n},t) \},
\end{eqnarray}
where $r_\mathrm{trap}$ is the trapping rate, assumed to be the same for chromophores 3 and 4. 
When comparing to Eq.~(\ref{heom}), it can be seen that adding this Liouvillian makes the population 
of chromophore 3 and 4 decay faster than for the other chromophores.

The density operator describing EET in the FMO complex can be written as 
\begin{eqnarray}
\label{rho}
\hat{\varrho}(t) = \sum_{k=1}^{7} \varrho_{kk} (t) \ket{k} \bra{k}
+ \sum_{k=1}^{7} \sum_{l>k}^{7}
\left( \varrho_{kl} (t) \ket{k} \bra{l} + \varrho_{kl}^{\ast} (t) \ket{l} \bra{k} \right),
\end{eqnarray}
where $\ket{k}$ are the site basis states defined in Eq.~(\ref{eq:sitebasis}). Here, $\varrho_{kk}$ 
is the population of an excitation at chrompohore $k$ and $\varrho_{kl}$ describes the coherence 
between chromophores $k$ and $l$. Note that $\hat{\varrho}$ can be viewed as a $7 \times 7$ 
dimensional Hermitian matrix due to the restriction to one coherent excitation at each instant of time.

\section*{\sffamily \Large MEASURE OF BIPARTITE QUANTUM CORRELATIONS} 
\label{sec:nonlocality}
Quantum nonlocality is a property of composite quantum systems whose subsystems show 
correlations that are too strong to be explained by a local realistic theory \cite{Einstein1935}, 
i.e., a theory where physical variables (such as, e.g., positions and momenta of particles) are 
assumed to have well-defined local values prior to measurement. The existence of quantum 
nonlocality was discovered when Bell derived \cite{Bell1964} an upper bound (Bell's inequality) for 
correlations between two systems to be local, and then showed that the correlations within certain 
composite quantum systems may exceed this limit.

Since Bell's original formulation of his inequality, there have been other Bell-like inequalities. 
One such that is suitable for investigating bipartite nonlocality for TLSs like the chromophores in 
the FMO complex is the Clauser-Horne-Shimony-Holt (CHSH) inequality \cite{Clauser1969}. 
It states that any set of variables $a, a',b$, and $b'$ that can take values $\pm 1$ must satisfy
\begin{eqnarray}
\left| \langle ab \rangle + \langle ab' \rangle +\langle a'b \rangle - \langle a'b' \rangle \right|
\leq 2
\label{eq:chsh}
\end{eqnarray}
if a local realistic theory applies to the pairs $a,a'$ and $b,b'$ at distant locations. Here, 
$\langle ab \rangle$ denotes the average of the product of the outcomes $a$ and $b$.

To test the validity of the CHSH inequality for measurements on two distant TLSs, each 
characterized by the Pauli operators $\boldsymbol{\hat{\sigma}} = (\hat{\sigma}_1,\hat{\sigma}_2,
\hat{\sigma}_3)$, described as a composite quantum system with density operator 
$\hat{\rho}$, the operator
\begin{eqnarray}
\hat{\mathfrak{B}}_\textrm{CHSH} & = &
({\bf a} \cdot \hat{\boldsymbol{\sigma}}) \otimes ({\bf b} \cdot\hat{\boldsymbol{\sigma}}) + 
({\bf a} \cdot \hat{\boldsymbol{\sigma}}) \otimes ({\bf b}' \cdot\hat{\boldsymbol{\sigma}}) 
\nonumber \\
 & & 
+ ({\bf a}' \cdot \hat{\boldsymbol{\sigma}}) \otimes ({\bf b} \cdot\hat{\boldsymbol{\sigma}}) - 
({\bf a}' \cdot \hat{\boldsymbol{\sigma}}) \otimes ({\bf b}' \cdot
\hat{\boldsymbol{\sigma}}) 
\label{CHSH-operator}
\end{eqnarray}
with ${\bf a},{\bf a}',{\bf b}$, and ${\bf b}'$ unit vectors in $R^3$, can be used, as the  
measurements of $({\bf a} \cdot \boldsymbol{\hat{\sigma}}), ({\bf a}' \cdot 
\boldsymbol{\hat{\sigma}}),({\bf b} \cdot \boldsymbol{\hat{\sigma}})$, and $({\bf b}' \cdot 
\boldsymbol{\hat{\sigma}})$ on the respective TLSs have outcomes $\pm 1$. Thus, by comparing 
with Eq.~(\ref{eq:chsh}), one may conclude that the correlation in $\hat{\rho}$ is nonlocal, i.e., 
cannot be accounted for by any local realistic theory, if there exists ${\bf a},{\bf a}',{\bf b}$, 
and ${\bf b}'$ such that $|\langle \hat{\mathfrak{B}}_\textrm{CHSH}\rangle | = |\Tr \left( \hat{\rho}
\hat{\mathfrak{B}}_\textrm{CHSH} \right) | > 2$.

A necessary and sufficient condition for the correlation between any two TLSs to be nonlocal 
has been developed by Horodecki {\it et al.} \cite{Horodecki1995}. This criterion is based on 
a quantity $M(\hat{\rho})$ that maximizes the expectation value of the Bell operator in  
Eq.~(\ref{CHSH-operator}) such that
\begin{eqnarray}
\max_{{\bf a},{\bf a}',{\bf b},{\bf b}'} |\langle \hat{\mathfrak{B}}_\textrm{CHSH}\rangle | =
2\sqrt{M(\hat{\rho})} .
\end{eqnarray}
$M(\hat{\rho})$ is found as
\begin{eqnarray}
M(\hat{\rho})=\mu_p + \mu_q,
\end{eqnarray}
where $\mu_p$ and $\mu_q$ are the two greatest eigenvalues of $\left| T(\hat{\rho}) \right|^2$, 
$T(\hat{\rho})$ being the correlation matrix 
\begin{eqnarray}
T(\hat{\rho}) =
\begin{pmatrix}
  t_{11} &  t_{12} &  t_{13} \\
  t_{21} &  t_{22} &  t_{23} \\
  t_{31} &  t_{32} &  t_{33}
 \end{pmatrix} 
\end{eqnarray}
with matrix elements $t_{\alpha\beta} = \Tr(\hat{\rho}\hat{\sigma}_{\alpha} \otimes 
\hat{\sigma}_{\beta})$ for all combinations of Pauli operators $\hat{\sigma}_{\alpha}$. The 
correlation is nonlocal whenever $1 < M(\hat{\rho}) \leq 2$, where the maximum value is given 
by the Cirel'son bound \cite{Cirelson1980} $|\langle \hat{\mathfrak{B}}_\textrm{CHSH}\rangle | 
\leq 2\sqrt{2}$. This motivates that the quantity \cite{Horst2013}
\begin{eqnarray}
B(\hat{\rho})=\sqrt{\max \{ (M(\hat{\rho}) - 1),0 \} } 
\label{eq:nonlocalitymeasure}
\end{eqnarray}
can be used as a measure of the amount of nonlocality. This Bell-CHSH measure is directly 
comparable with concurrence, defined as \cite{Wootters1998}  
\begin{eqnarray}
C(\hat{\rho}) = \max \{ \lambda_1 - \lambda_2 - \lambda_3 - \lambda_4 , 0 \} , 
\end{eqnarray}
where $\lambda_1, \ldots \lambda_4$ are the decreasingly ordered eigenvalues of the positive 
operator 
\begin{eqnarray} 
\hat{R} = \sqrt{\sqrt{\hat{\rho}} \sigma_2 \otimes \sigma_2 \hat{\rho}^{\ast} \sigma_2 \otimes 
\sigma_2 \sqrt{\hat{\rho}}} 
\end{eqnarray}
with complex conjugation taken in the computational product basis. Concurrence uniquely 
determines the entanglement of formation \cite{bennett1996} of two TLSs; as such, $C(\hat{\rho})$ 
is an entanglement measure in its own right. It has been used to analyze the amount of bipartite 
entanglement in the FMO complex by Sarovar {\it et al.} \cite{Sarovar2010}.

The time evolution of the composite system of  chromophores $\mathit{n}$ and $\mathit{m}$ in 
the FMO complex can be found by calculating the reduced density operator from the full density 
operator $\hat{\varrho}$ given in Eq.~(\ref{rho}). This is done by tracing $\hat{\varrho}$ 
over the other five chromophore degrees of freedom. The resulting reduced density operator 
takes the form
\begin{eqnarray}
\nonumber
\hat{\rho}_{mn} = \Tr_{r\neq n,m} \hat{\varrho} & = & 
\sum_{k,l,p,q=0,1} \rho_{kl;pq}^{mn} \ket{S_{k}^{m} S_{l}^{n}} \bra{S_{p}^{m}  S_{q}^{n}}
\nonumber \\ 
 & = & [\Tr (\hat{\varrho})-(\varrho_{mm} + \varrho_{nn})] \ket{S_{0}^{m} S_{0}^{n}} 
\bra{S_{0}^{m}  S_{0}^{n}} 
\nonumber \\
 & & + \varrho_{mm} \ket{S_{0}^{m} S_{1}^{n}} \bra{S_{0}^{m} S_{1}^{n}}  + 
\varrho_{mn} \ket{S_{0}^{m} S_{1}^{n}} \bra{S_{1}^{m} S_{0}^{n}} 
\nonumber \\
 & & + \varrho_{mn}^{\ast} \ket{S_{1}^{m} S_{0}^{n}} \bra{S_{0}^{m} S_{1}^{n}} + 
\varrho_{nn}\ket{S_{1}^{m} S_{0}^{n}} \bra{S_{1}^{m} S_{0}^{n}} ,  
\label{eq:reducedstate}
\end{eqnarray}
where we have taken into account that HEOM and excitation trapping are not trace preserving, 
i.e., that $\Tr (\hat{\varrho}) \leq 1$. Note that $\hat{\rho}_{mn}$ can be viewed as a 
$4\times 4$ Hermitian matrix.  One finds  
\begin{eqnarray}
\mu_1=\mu_2=4\left| \varrho_{mn} \right|^2, \ \ 
\mu_{3} = [\Tr (\hat{\varrho})-2(\varrho_{mm} + \varrho_{nn})]^2 , 
\label{eq:eigenvalues}
\end{eqnarray}
which implies 
\begin{eqnarray}
B(\hat{\rho}_{mn}) = \sqrt{\max \left\{ \left( \max\{8\left| \varrho_{mn} \right|^2,4\left| \varrho_{mn} \right|^2 
+[\textrm{Tr} (\hat{\varrho})-2(\varrho_{mm} + \varrho_{nn})]^2\} - 1\right) , 0 \right\} } .
\label{eq:Mmatrix}
\end{eqnarray}
Similarly, concurrence for this class of states takes the form 
\begin{eqnarray}
C(\hat{\rho}_{mn}) = 2\left| \varrho_{mn} \right| , 
\end{eqnarray}
which coincides with the square root of $\mu_1$. 

Recent attempts to quantify the amount of coherence in quantum states \cite{aberg2006,baumgratz2014} 
have lead to different types of coherence measures. One of these is the $l_1$ norm measure of 
coherence \cite{baumgratz2014}, defined as 
\begin{eqnarray}
\mathcal{C}_{l_1} (\hat{\rho}) = \sum_{k\neq l} \left| \rho_{kl} \right| ,  
\end{eqnarray}
i.e., the sum of the absolute value of the off-diagonal elements of the density operator. 
For $\hat{\rho}_{mn}$ this reduces to  
\begin{eqnarray}
\mathcal{C}_{l_1} (\hat{\rho}_{mn}) = 2\left| \varrho_{mn} \right| ,  
\end{eqnarray}
which implies that concurrence and the $l_1$ norm measure of coherence are in fact identical 
quantities for all chromophore pairs. Thus, a nonzero concurrence can equally well be interpreted 
as an expression for having a nonzero coherence, rather than being a sign of correlation. This 
is a consequence of the restriction to the single excitation subspace of the Hamiltonian in 
Eq.~(\ref{hamiltonian}). Thus, the restriction itself implies the existence of entanglement as 
a consequence of the nonvanishing coherence in the system. 

Nonlocality, on the other hand, is essentially different from coherence as it also depends on 
the populations $\varrho_{mm}$ and $\varrho_{nn}$ of the chromophores via the 
nondegenerate eigenvalue $\mu_3$ of $|T(\hat{\rho})|^2$. This makes the Bell-CHSH 
measure a proper quantifier of correlation between chromophore pairs.

We may ask under what circumstances a chromophore pair exhibiting nonvanishing coherence 
is nonlocally correlated. As follows from the explicit expressions in Eq.~(\ref{eq:Mmatrix}), the 
nonlocality condition relies on the relation between the populations $\varrho_{mm}$ and 
$\varrho_{nn}$ of the pair {\it and} the coherence $2\left| \varrho_{mn} \right|$. Since the 
latter is bounded by the former as \cite{Leon-Montiel2015} 
\begin{eqnarray}
\left| \varrho_{mn} \right| \leq \sqrt{\varrho_{mm} \varrho_{nn}} ,
\label{eq:positivity}
\end{eqnarray} 
which follows from positivity of the reduced density matrix, we see that $\mu_1$ and $\mu_2$ 
become very small unless the system is considerably localized to the $(m,n)$ pair, in case 
of which $\mu_3$ also becomes large. Alternatively, $\mu_3$ can be close to the critical 
value $1$ sufficient for nonlocal correlations if $\varrho_{mm}$ and $\varrho_{nn}$ 
are both very small (in the order of a few $\%$, say) and $\Tr (\hat{\varrho}) 
\lesssim 1$; however, this cannot give rise to any nonlocality. To see this, we first 
note that 
\begin{eqnarray}
M(\rho_{mn}) & = & \mu_1 + \mu_3 \leq 
4\varrho_{mm} \varrho_{nn} + \left( \Tr (\hat{\varrho}) - 2\varrho_{mm} - 2\varrho_{nn} \right)^2 
\nonumber \\ 
 & = & \Tr (\hat{\varrho}) \left( \Tr (\hat{\varrho})  - 4\varrho_{mm} -4\varrho_{nn} \right) + 
4\varrho_{mm}^2 + 4\varrho_{nn}^2 + 12 \varrho_{mm}\varrho_{nn} 
\nonumber \\ 
 & \lesssim & 1  - 4\left( \varrho_{mm} + \varrho_{nn}  - \varrho_{mm}^2 - \varrho_{nn}^2 
- 3 \varrho_{mm}\varrho_{nn} \right) 
\end{eqnarray}
by combining Eqs.~(\ref{eq:eigenvalues}), (\ref{eq:positivity}), and $\Tr (\hat{\varrho}) 
\lesssim 1$. It is straightforward to see that $\varrho_{mm} + \varrho_{nn}  - \varrho_{mm}^2 - 
\varrho_{nn}^2 - 3 \varrho_{mm}\varrho_{nn} \geq 0$ for small $\varrho_{mm}$ and 
$\varrho_{nn}$, which implies that 
\begin{eqnarray}
M(\rho_{mn}) \leq 1 
\end{eqnarray} 
excluding the possibility of nonlocal correlations in this case.
We conclude that only chromphore pairs for which the population is large can be nonlocally correlated.

\section*{\sffamily \Large NUMERICAL DETAILS}
\label{sec:numerical}
The parameters of our numerical model are chosen in accordance with previous work on 
EET in the FMO complex. As the electronic FMO Hamiltonian for \textit{Chlorobaculum tepidum} 
in the chromophore site basis, we use \cite{Adolphs2006}
\begin{eqnarray}
H_e = \left( \begin{array}{ccccccc}
200 & -87.7 & 5.5 & -5.9 & 6.7 & -13.7 & -9.9 \\
-87.7 & 320 & 30.8 & 8.2 & 0.7 & 11.8 & 4.3 \\
5.5 & 30.8 & 0 & -53.5 & -2.2 & -9.6 & 6.0 \\
-5.9 & 8.2 & -53.5 & 110 & -70.7 & -17.0 & -63.3 \\
6.7 & 0.7 & -2.2 & -70.7 & 270 & 81.1 & -1.3 \\
-13.7 & 11.8 & -9.6 & -17.0 & 81.1 & 420 & 39.7 \\
-9.9 & 4.3 & 6.0 & -63.3 & -1.3 & 39.7 & 230 \\
\end{array} \right) ,
\label{eq:hmatrix}
\end{eqnarray} 
where all numbers are in units of cm$^{-1}$ with a total offset of 12 210 cm$^{-1}$. The 
reorganization energy $\lambda_k$ and the relaxation time-scale $\gamma_k$ are assumed 
to have the same values, 35 cm$^{-1}$ and 50 fs$^{-1}$ \cite{Read2008}, respectively, for all 
seven chromophores \cite{Ishizaki2009a}. The time scale $r_\mathrm{trap}^{-1}$ for the 
trapping by the reaction center is set to 1 ps \cite{Cho2005, Adolphs2006} and the bath 
temperature to 300 K (same as in Ref. \cite{Ishizaki2009a}).

\begin{figure}[ht]
\centering
\includegraphics[width=0.6\textwidth]{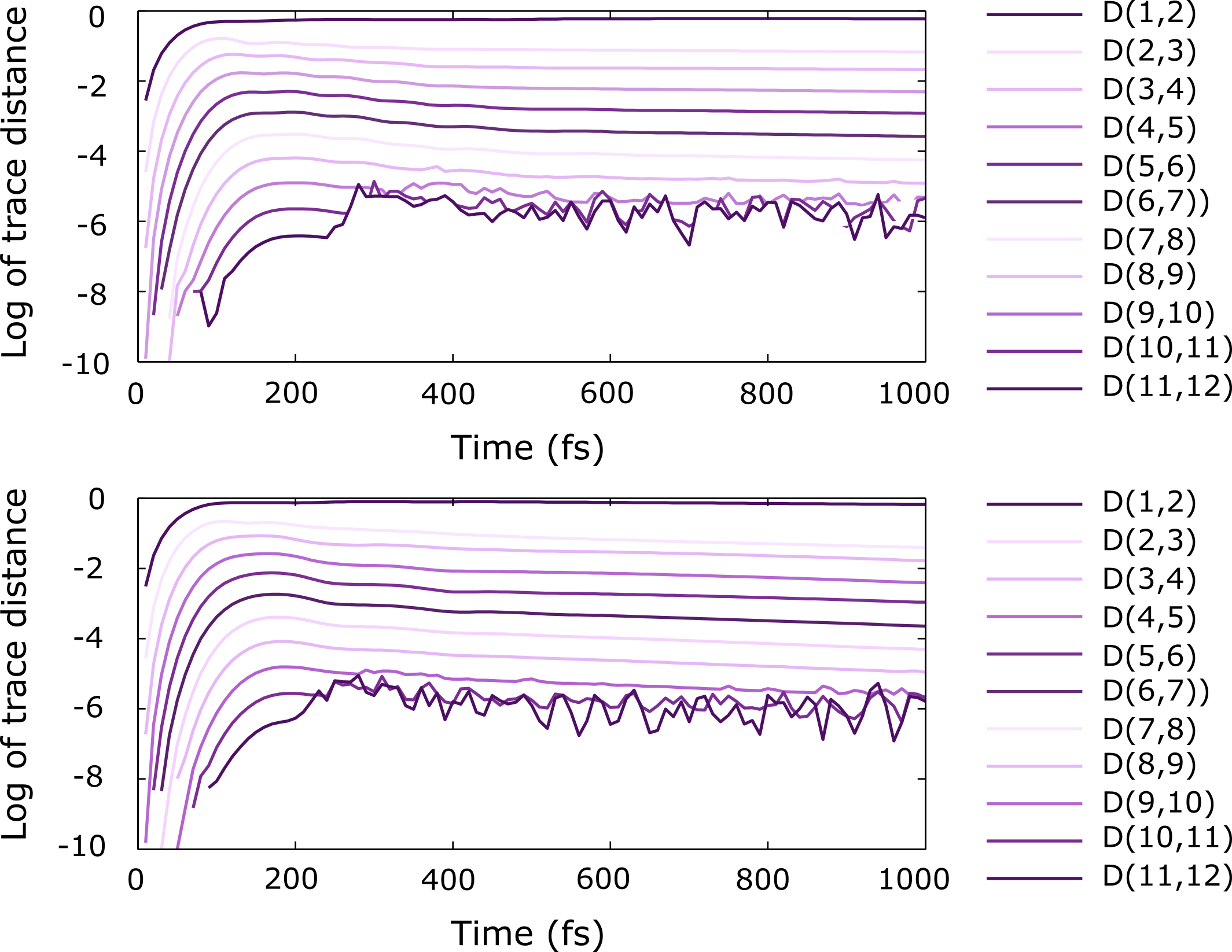}
\caption{Convergence of the density operator with respect to the truncation level of the HEOM. The 
change in the density operator when going from level $N$ to level $N+1$ is characterized 
by the logarithm of the corresponding trace distance $\textrm{D}(N,N+1) \equiv D(\hat{\varrho}_N,
\hat{\varrho}_{N+1}) = \frac{1}{2} \Tr \left| \hat{\varrho}_N -\hat{\varrho}_{N+1}  \right|$. 
The initial excitation is either on chromophore 1 (upper panel) or chromophore 6 (lower panel). 
Notice how at the higher truncation levels ($N\gtrsim 10$) the trace distance is comparable 
to the numerical noise.}
\label{Convergence}
\end{figure}

The HEOM given by Eq.~(\ref{heom}) is numerically integrated in the range 0 to 1000 fs using the
Runge-Kutta-Dormand-Prince method \cite{Dormand1980}. To measure the convergence of 
the HEOM solution, we use the trace distance \cite{Nielsen2000}
\begin{eqnarray}
D(\hat{\varrho}_N,\hat{\varrho}_{N+1}) = 
\frac{1}{2} \Tr \left| \hat{\varrho}_N -\hat{\varrho}_{N+1}  \right|
\end{eqnarray}
for density operators $\hat{\varrho}_N$ and $\hat{\varrho}_{N+1}$ at truncation level $N$ and 
$N+1$, respectively. In our simulations 
in the next section, we use a truncation level $N=12$, which implies an accuracy in the order 
of $10^{-5}$, as can be seen from Fig.~\ref{Convergence}. 

\section*{\sffamily \Large RESULTS AND DISCUSSION}
\label{sec:results}
\begin{figure}[ht]
\centering
\includegraphics[width=0.6\textwidth]{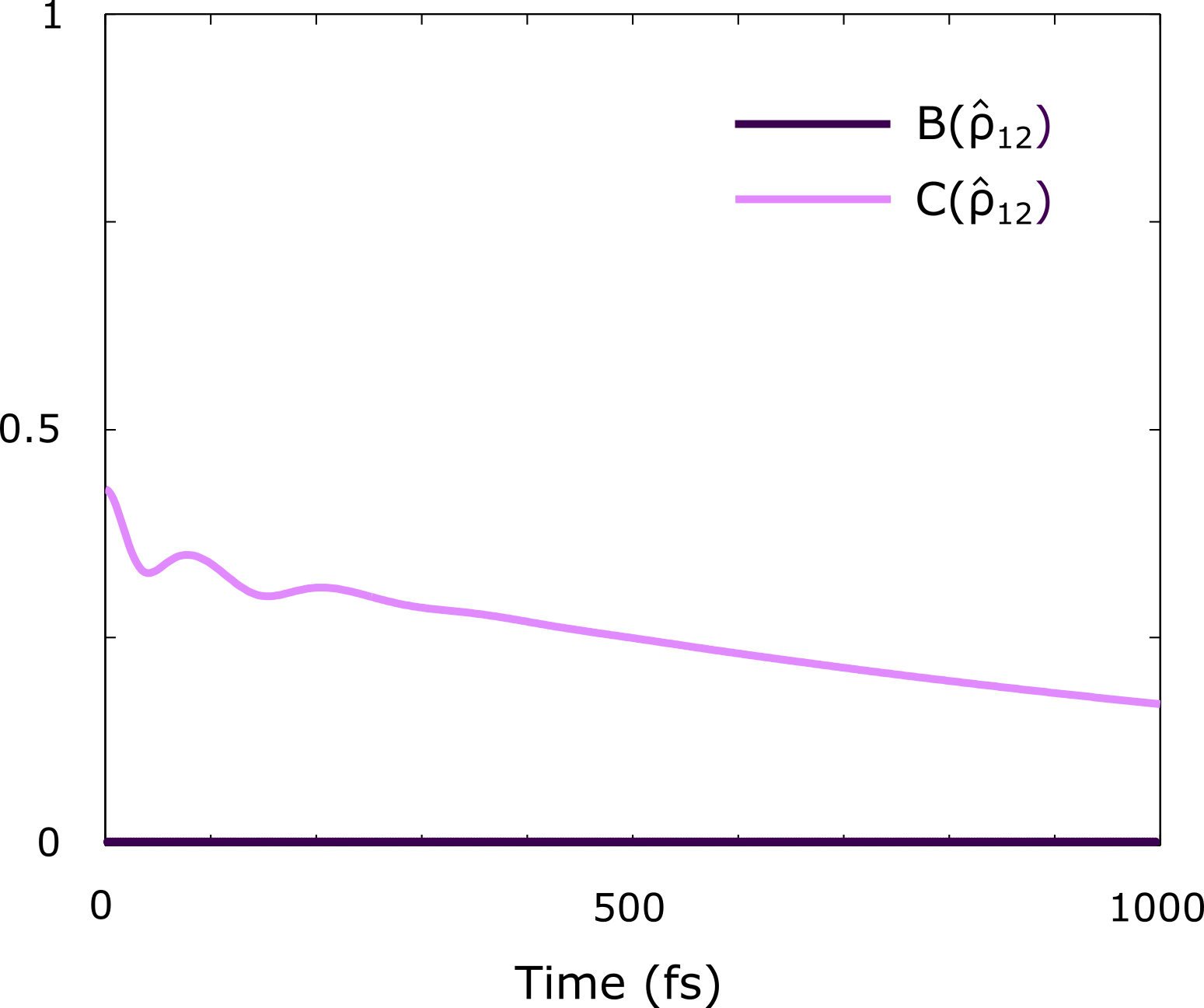}
\caption{Nonlocality $B(\hat{\rho}_{12})$ and entanglement $C(\hat{\rho}_{12})$ 
for the chromophore pair 1 and 2 as a function of time for FRET initial state when 
exciton states projected on chromophore 1 is used as initial condition.
Note that the nonlocality $B(\hat{\rho}_{12})$ vanishes at all times.}
\label{IC1_12_ny}
\end{figure}

\begin{figure}[ht]
\centering
\includegraphics[width=0.6\textwidth]{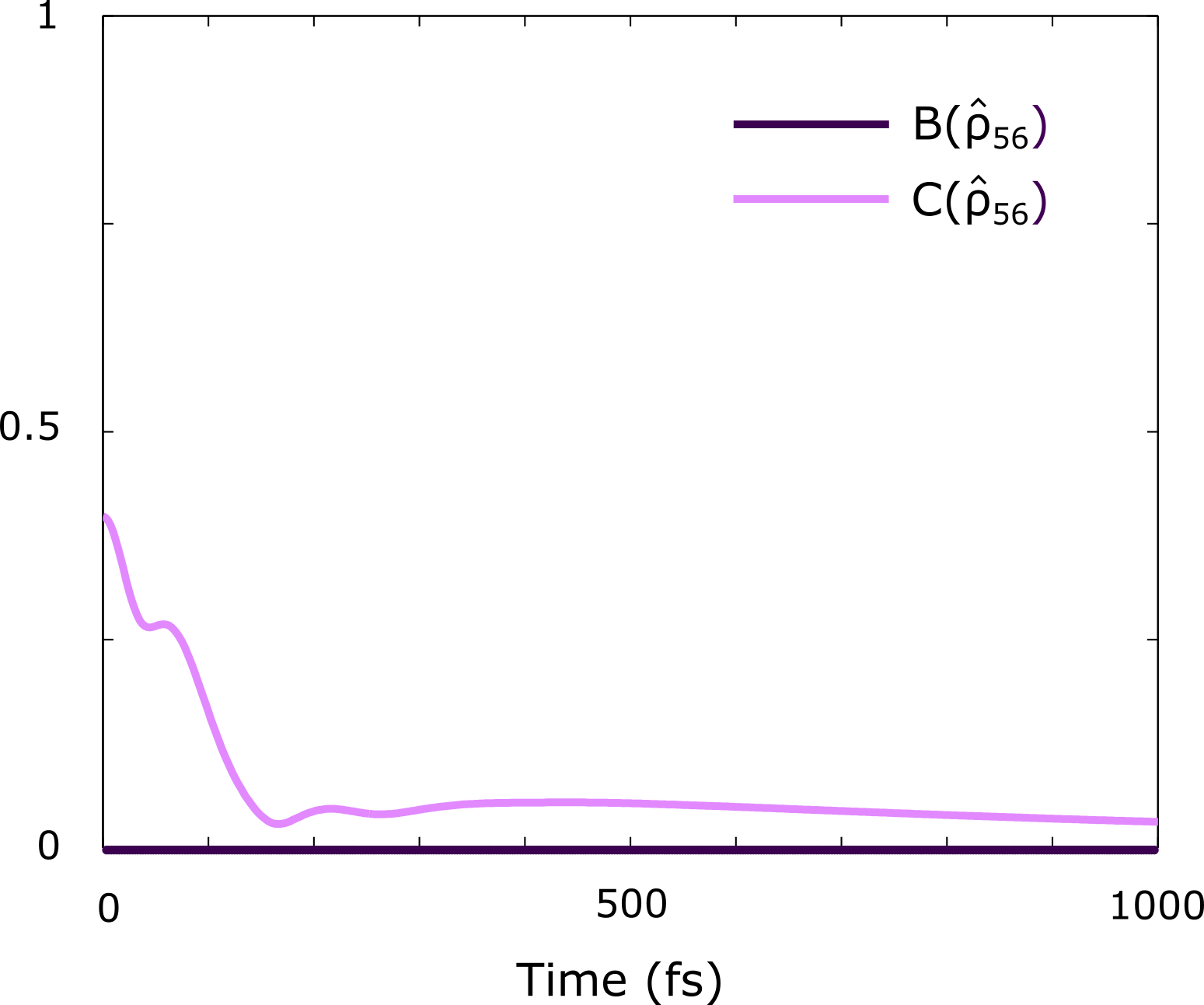}
\caption{Nonlocality $B(\hat{\rho}_{56})$  and entanglement $C(\hat{\rho}_{56})$ 
for the chromophore pair 5 and 6 as a function of time for FRET initial state when 
exciton states projected on chromophore 6 is used as initial condition. Note that the nonlocality 
$B(\hat{\rho}_{12})$ vanishes at all times.}
\label{IC6_56_ny}
\end{figure}

\begin{figure}[ht]
\centering
\includegraphics[width=0.6\textwidth]{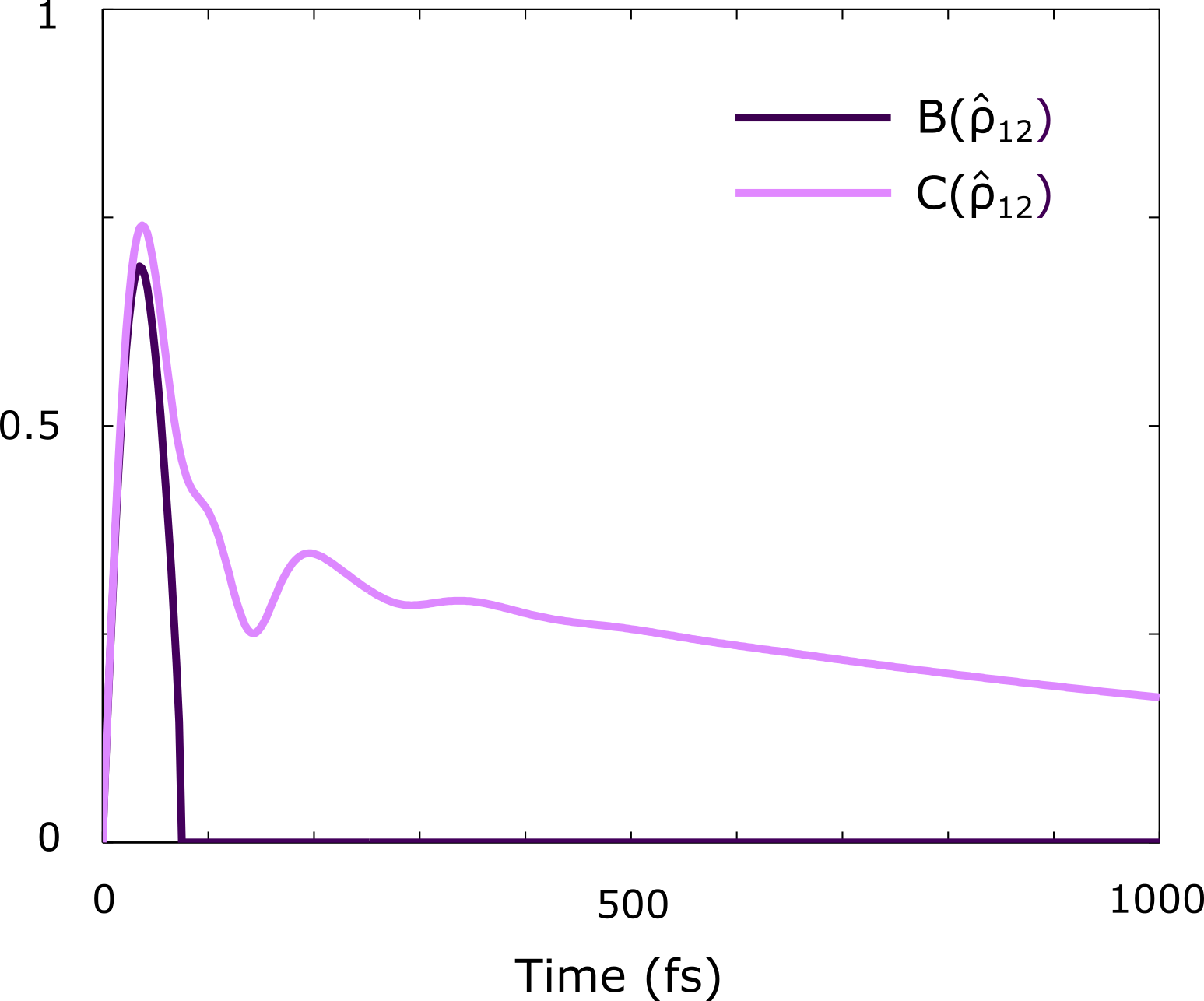}
\caption{Nonlocality $B(\hat{\rho}_{12})$ and entanglement $C(\hat{\rho}_{12})$ 
for the chromophore pair 1 and 2 as a function of time when chromophore 1 is 
initially excited. Note how the nonlocality drops to zero within about 80 fs.}
\label{IC1_12}
\end{figure}

\begin{figure}[ht]
\centering
\includegraphics[width=0.6\textwidth]{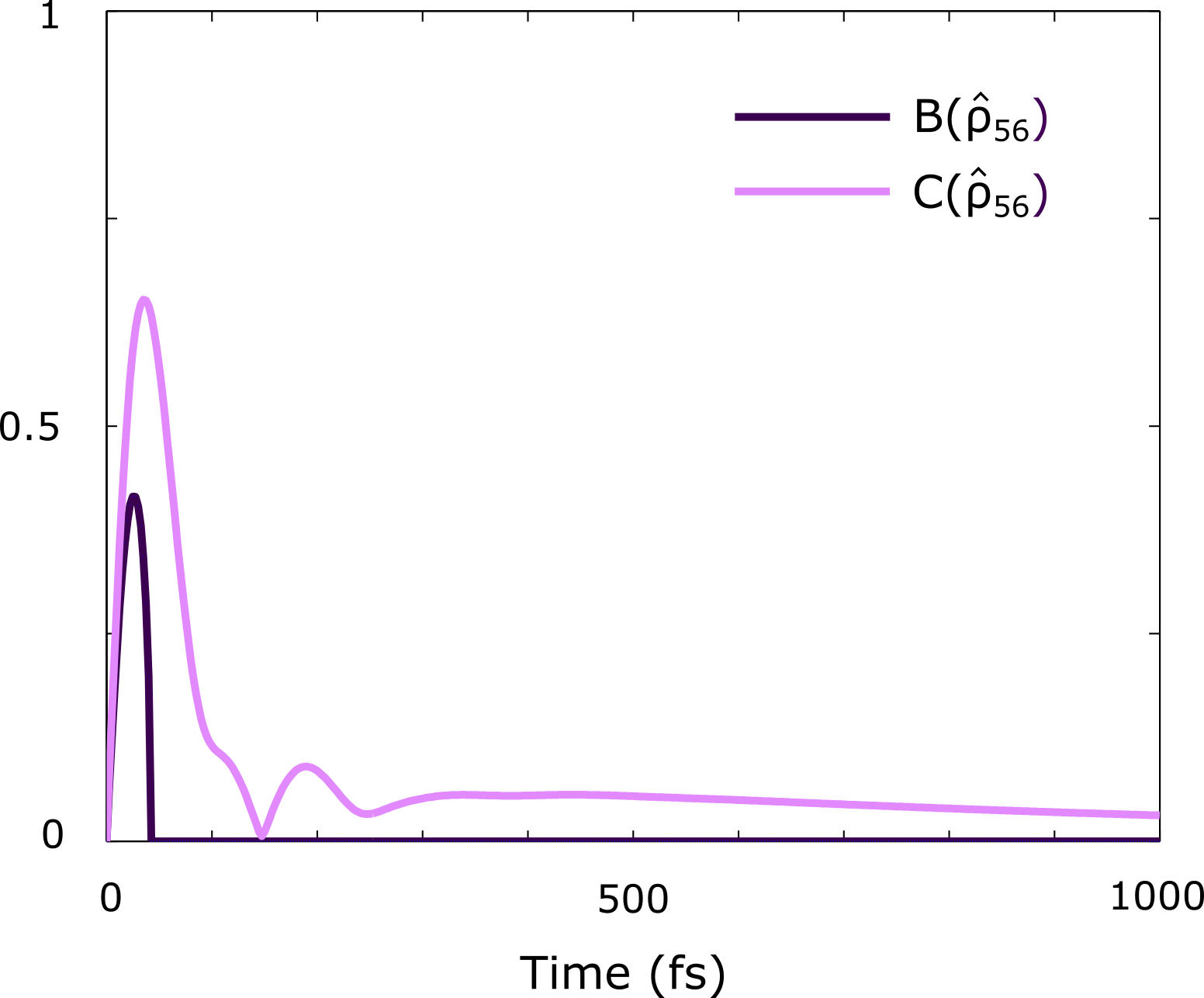}
\caption{Nonlocality $B(\hat{\rho}_{56})$ and entanglement $C(\hat{\rho}_{56})$ 
for the chromophore pair 5 and 6 as a function of time when chromophore 6 is 
initially excited. Note how the nonlocality drops to zero within about 40 fs.}
\label{IC6_56}
\end{figure}

Our simulations show that no nonlocality is found when the initial states are given by 
Eq.~(\ref{ic_fret}), corresponding to FRET from the antenna molecules to the FMO, although 
entanglement still exists. In particular, a considerable amount of quantum entanglement is 
found for chromophore 1 and 2 when the exciton states are projected on chromophore 1 
as well as for chromophore 5 and 6 when the exciton states are projected on chromophore 6. 
These results can be seen in Figs.~\ref{IC1_12_ny} and \ref{IC6_56_ny}. On the other hand, 
bipartite quantum nonlocality exists when the localized initial conditions according to 
Eq.~(\ref{ic_localized}) are used. For these initial conditions, nonlocality is found for two 
chromophore pairs; chromophore 1 and 2 when chromophore 1 is initially excited, and 
chromophore 5 and 6 when chromophore 6 is initially excited. These results are presented 
in Figs.~\ref{IC1_12} and \ref{IC6_56} together with the time dependence of the bipartite 
entanglement for the same pairs of chromophores. 

To get futher insights into these findings we analyze in more detail the structure of the 
evolution arising from the two types of initial conditions. Let us start with the FRET case, 
where the initial states are given by Eq.~(\ref{ic_fret}). By writing the exciton states 
$\ket{e_r} = \sum_{k=1}^7 c_{rk} \ket{k}$, we find  
\begin{eqnarray}
\hat{\varrho}_{\textrm{FRET}}^{x} & = & \sum_{r=1}^7 c_{rx}^2 \ket{e_r} \bra{e_r} ,  
\end{eqnarray}
where we have used that all $c_{rk}$ are real-valued and we have ordered $\ket{e_r}$ with 
increasing energy. The reduced density matrix for the $(m,n)$ pair is characterized by the 
site probablities $\varrho_{mm},\varrho_{nn}$, and the off-diagonal term $\varrho_{mn}$ 
given by  
\begin{eqnarray}
\varrho_{mm} & = & \sum_{r=1}^7 c_{rx}^2 c_{rm}^2 , 
\nonumber \\ 
\varrho_{nn} & = & \sum_{r=1}^7 c_{rx}^2 c_{rn}^2 ,  
\nonumber \\ 
\varrho_{mn} & = & \sum_{r=1}^7 c_{rx}^2 c_{rm} c_{rn} . 
\end{eqnarray} 

We focus on the $x=1$ case. As demonstrated above, only chromphore pairs for which the 
population is large can exhibit nonlocal correlations; thus, we only need to consider the 
dominant terms in the density operator of the FMO complex. By inspection of the explicit 
eigenvectors, we find that $\hat{\varrho}_{\textrm{FRET}}^{x=1}$ is dominated by the pure 
state components $\ket{e_3}$ and $\ket{e_6}$. In fact, $c_{31}^2 \approx 0.769$ and $c_{61}^2 
\approx 0.208$; thus, these two exciton states populate almost $98 \%$ of this initial state and 
we can safely ignore all the other exciton states. We further find that both $\ket{e_3}$ 
and $\ket{e_6}$ are to a large extent localized to the first two chromophores, as can be seen 
from the expansion coefficients 
\begin{eqnarray}
c_{31} & = & 0.877, \ \ c_{32} = 0.440 , 
\nonumber \\ 
c_{61} & = & -0.456, \ \ c_{62} = 0.871 ,  
\end{eqnarray}
obtained by numerical diagonalization of the electronic Hamiltonian $H_e$. Thus, correlation 
is essentially concentrated to the first two chromophores, and can be expressed in terms of the 
matrix elements 
\begin{eqnarray} 
\varrho_{12} & \approx & c_{31}^3 c_{32} + c_{61}^3 c_{62} 
\nonumber \\ 
 & \equiv &  c_{31}^2 \left| c_{31} c_{32} \right| - c_{61}^2 \left| c_{61} c_{62} \right| = 0.214 , 
\nonumber \\
\varrho_{11} & \approx & c_{31}^4 + c_{61}^4 = 0.635 ,  
\nonumber \\
\varrho_{22} & \approx & c_{31}^2 c_{32}^2 + c_{61}^2 c_{62}^2 = 0.307 . 
\label{eq:approximate}
\end{eqnarray}
These numerical values imply 
\begin{eqnarray}
C(\hat{\rho}_{12}) & = & 2\left| \varrho_{12} \right| \approx 0.428 ,   
\nonumber \\ 
\mu_1 & = & \mu_2 = \left[ C (\hat{\rho}_{12}) \right]^2 \approx 0.183, 
\nonumber \\ 
\mu_3 & = & (1-2\varrho_{11} -2\varrho_{22})^2 \approx 0.781 ;  
\end{eqnarray}
thus, $\hat{\varrho}_{\textrm{FRET}}^{x=1}$ shows entanglement ($C>0$) but no nonlocality 
($B=0$ since $\mu_p + \mu_q = 0.183 + 0.781 = 0.964 < 1$) between the first two 
chromophores. Since correlations essentially are concentrated to this chromophore-pair, it follows 
that all chromophore pairs are locally correlated at $t=0$, given the FRET initial condition. 
 
On the other hand, we note that the two dominating exciton states $\hat{\varrho}^{e_3} = 
\ket{e_3} \bra{e_3}$ and $\hat{\varrho}^{e_6} = \ket{e_6} \bra{e_6}$
are strongly nonlocal between chromophores $1$ and $2$; indeed, one finds 
\begin{eqnarray}
B(\hat{\rho}_{12}^{e_3}) & = & C(\hat{\rho}_{12}^{e_3}) \approx 2|c_{31}c_{32}| = 0.801, 
\nonumber \\ 
B(\hat{\rho}_{12}^{e_6}) & = & C(\hat{\rho}_{12}^{e_6}) \approx 2|c_{61}c_{62}| = 0.822 ,   
\end{eqnarray}
where we have used that $B=C$ for pure states of any TLS-pair \cite{Horst2013}. By comparing 
these expressions with the expression for $\varrho_{12}$ in Eq.~(\ref{eq:approximate}), we find 
\begin{eqnarray} 
\varrho_{12} & \approx & c_{31}^2 B(\hat{\rho}_{12}^{e_3}) - c_{61}^2 B(\hat{\rho}_{12}^{e_6}) ,   
\end{eqnarray} 
which explicitly entails that the essential source for the disappearance of nonlocality when 
mixing the exicton state is the destructive quantum-mechanical interference (relative minus sign) 
between the first-site components $c_{31}$ and $c_{61}$ of $\ket{e_3}$ and $\ket{e_6}$. 

A similar analysis of the $x=6$ case can be carried out with the same conclusion that  
all chromophore pairs are locally correlated, but entangled at $t=0$.

FRET are stationary states under the action of the electronic Hamiltonian, but they may undergo 
nontrivial evolution under influence of the environment. Thus, the environment is a potential source 
of nonlocal correlations to show up at $t>0$. However, as can be seen in Figs.~\ref{IC1_12_ny} 
and \ref{IC6_56_ny}, it turns out that nonlocal correlations never appear given the FRET initial 
conditions, despite the fact that entanglement is typically nonvanishing. 

Next, we consider the case of localized initial states $\varrho_{\textrm{localized}}^{x} = 
\ket{x} \bra{x}$, $x=1,6$. These states are not eigenstates of $H_e$ and can therefore in 
principle give rise to nonlocal correlations during time evolution. 

Let us determine for which chromophore pair(s) nonlocality can occur. In the short time limit, 
we may expect that ${\bf n} = 0$ is the dominant term in the HEOM. Thus, we may assume 
that $\hat{\varrho} (\delta t) \approx e^{-i\delta t \hat{H}_e} \hat{\varrho}_{\textrm{localized}}^x 
e^{i\delta t \hat{H}_e}$ for small $\delta t$ (from now on, we put $\hbar = 1$ for notational 
simplicity). By expanding in $\delta t$, we find to leading orders:  
\begin{eqnarray}
C(\hat{\rho}_{xn}) & \approx & 2\delta t \left| J_{xn} \right| , 
\nonumber \\ 
C(\hat{\rho}_{mn}) & \approx & 2\delta t^2 \left| J_{mx} J_{xn} \right| , 
\nonumber \\ 
B(\hat{\rho}_{xn}) & \approx & 
2 \delta t \sqrt{\max \left\{ \left( J_{xn}^2 - \sum_{l\neq n} J_{xl}^2 \right),0 \right\}} , 
\nonumber \\  
B(\hat{\rho}_{mn}) & = & 0, 
\label{eq:shorttime}
\end{eqnarray}
where $m,n \neq x$. Here, $J_{kl}$ are the real-valued dipole coupling parameters given by 
the off-diagonal terms of $H_e$. We may therefore draw the following conclusions about 
the short-time behavior of a localized initial state:  
\begin{itemize} 
\item[(i)] entanglement between the initially excited chromophore $x$ and any other chromophore 
$n$ increases linearly with $\delta t$ with proportionality factor given by the corresponding 
dipole coupling parameter $\left| J_{xn} \right|$; 
\item[(ii)] entanglement between any other pair of chromophores increases quadratically with 
$\delta t$; 
\item[(iii)] nonlocal correlation between chromophore $x$ and chromophore $n$ increases 
linearly with $\delta t$ provided
\begin{eqnarray}
J_{xn}^2 > \sum_{l\neq n} J_{xl}^2; 
\label{eq:inequality}
\end{eqnarray}
\item[(iv)] nonlocal correlations between any other pair of chromophores vanish. 
\end{itemize}

The inequality in Eq.~(\ref{eq:inequality}) can be satisfied for at most one pair. The explicit 
form of $H_e$ entails that only chromophores 1 and 2 can show nonlocal correlation after 
excitation of chromophore 1, while only chromophores 5 and 6 can show nonlocal correlation 
after excitation of chromophore 6. For these pairs we further notice that entanglement grows 
linearly with almost the same speed since $\left| J_{12} \right| \sim \left| J_{56} \right|$. 
As can be seen in Figs.~\ref{IC1_12} and \ref{IC6_56}, these results are confirmed in our 
numerical simulations of the HEOM, where we find nonlocality for $t$ less than $\sim 80$ fs 
and $\sim 50$ fs for $x=1$ and $x=6$, respectively. Indeed, at these instances of time, 
nonlocality suddenly disappears, although entanglement persists. We call this phenomenon 
{\it nonlocality sudden death}, being a striking manifestation of that nonlocality and entanglement 
are different properties for mixed quantum states \cite{Werner1989}. 

The absence of nonlocal correlation after its sudden death for localized initial states is a 
consequence of open system effects. In contrast, the complete absence of nonlocality in 
the FRET case is a consequence of the local nature of the initial state. Thus, we may 
conclude that the disappearance of nonlocality has different origin for the two types of 
initial conditions. 

\begin{figure}[ht]
\centering
\includegraphics[width=0.6\textwidth]{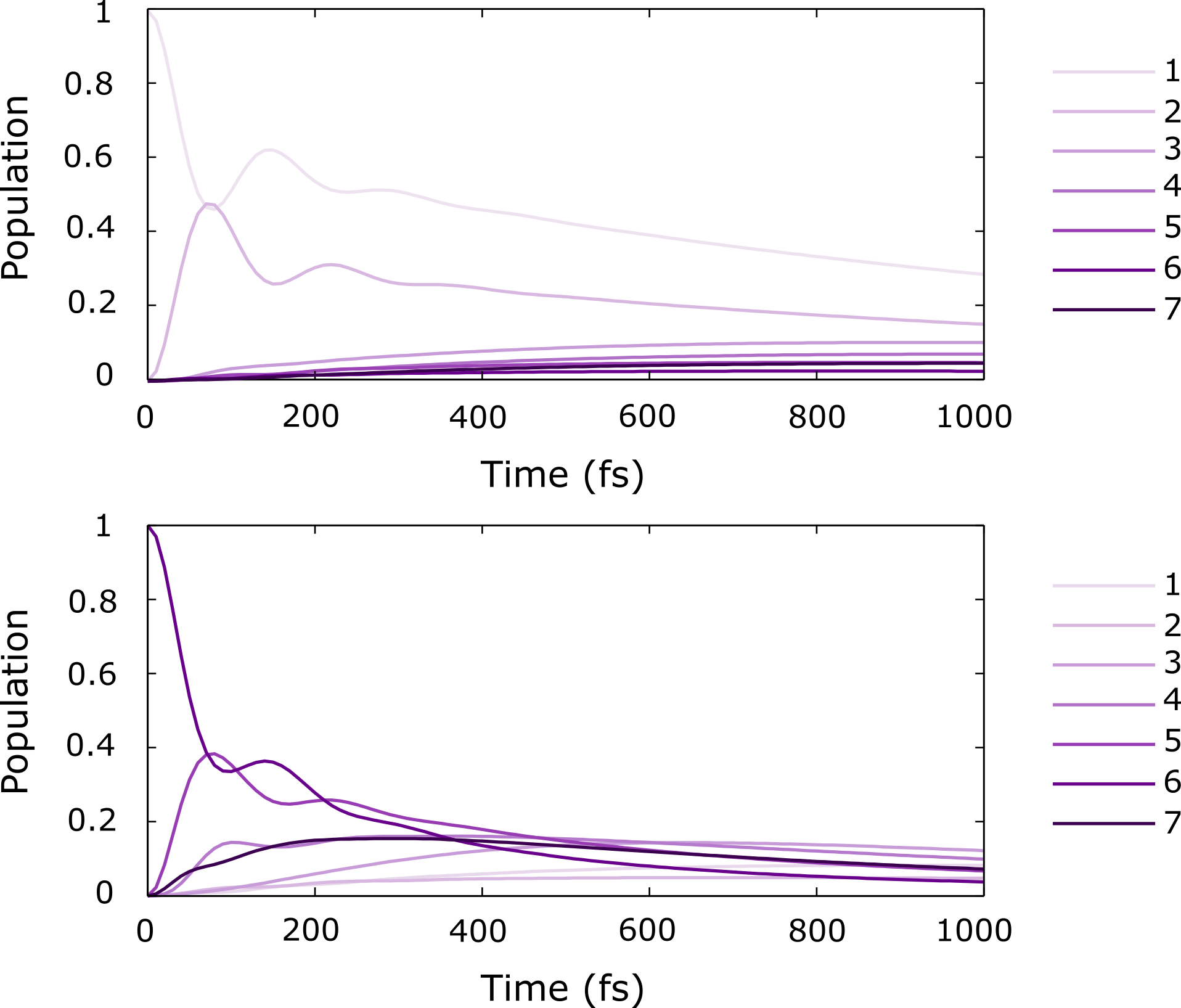}
\caption{Population of chromophores as a function of time when
chromophore 1 (upper panel) or chromophore 6 (lower panel) is initially excited. The 
population of the chromophores are given by the diagonal elements 
$\varrho_{11}, \ldots ,\varrho_{77}$ of the density operator in Eq.~(\ref{rho}).}
\label{Populations}
\end{figure}

As can be seen in Figs.~\ref{IC1_12_ny}-\ref{IC6_56}, bipartite nonlocality and entanglement 
behave very differently in EET in the FMO complex, when the HEOM method is used to 
model the dynamics. For the FRET initial conditions, nonlocality is not found for any pairs 
of chromophores, while entanglement is nonvanishing. Furthermore, for the localized initial 
conditions, entanglement dynamics is similar to the chromophores population dynamics shown 
in Fig.~\ref{Populations}, i.e., oscillating and exponentially decaying, while nonlocality 
does not show any oscillating features at all. Instead it drops to zero on a very short time 
scale, i.e., nonlocality may have a finite life-time despite the fact that entanglement 
undergoes an exponential decay. This nonlocality sudden death is a phenomenon analogous 
to entanglement sudden death, i.e., the occurrence of finite-time entanglement in coherent 
systems, which has been predicted \cite{Yu2004} and observed in quantum optical systems 
\cite{Almeida2007}. Similarly, nonlocality sudden death has very recently been demonstrated 
in a two-photon experiment \cite{liu2016}.

\section*{\sffamily \Large CONCLUSIONS}
\label{sec:conclusions}
This work has contributed to the investigation of EET in the FMO complex in considering
bipartite quantum nonlocality between different chromophores in addition to the bipartite 
quantum entanglement considered in previous studies \cite{Sarovar2010}. While entanglement 
can alternatively be interpreted as measuring the coherences (in the sense of $l_1$ norm of 
coherence) in the FMO complex, we have shown that nonlocality cannot be given this interpretation, 
but should instead be regarded a proper quantifier of quantum correlations between chromophore 
pairs. The numerical simulations show that nonlocality only exists for localized initial conditions. 
However, it is only observed for those chromophore pairs where one of them receives the 
initial excitation of the system and it drops to zero on a very short time scale (less than 100 fs). 
It should be noted that the behavior of nonlocality found in our simulations is related to 
the fact that the FMO complex can typically exhibit only one excitation at a time; by artificially 
including more simultaneous excitations would potentially induce a much more complicated 
nonlocality pattern accompanying the EET. The restriction to the single-excitation subspace 
would likely prevent nonlocal multipartite correlations \cite{zukowski02} among more than 
two chromophores, just as the absence of genuine multipartite entanglement for this class 
of states, well-known from the work by Coffmann {\it et al.} \cite{coffmann00}.

The fact that no nonlocality is observed when more realistic initial conditions (FRET from the 
antenna molecules to the FMO) are used indicates that nonlocality is of no importance when 
considering EET in the FMO in its natural habitat. The entanglement still present, represents 
local correlations in the sense that they can be described using a theory incorporating local 
realism \cite{Werner1989}. Hence, the correlations between pairs of chromophores can be 
either quantum or classical, i.e., it cannot be ruled out that EET in the FMO complex just as 
well could be explained from an underlying local realistic framework. Whether this is a result 
of the model used in this study or actually is according to the laws of nature, remains an 
open question. In relation to this latter remark, we further note that to experimentally examine 
the behavior of nonlocality in the FMO complex would be very challenging because it would 
require the ability to perform local measurements in at least two different bases, which would be 
hard to achieve due to the short time scale and the small distances between the chromophores. 

Finally, the occurence of nonlocality sudden death found in our simulations with localized initial 
conditions is another feature that seems to indicate that persistent quantum nonlocality is probably 
rare in biological systems. 

\subsection*{\sffamily \Large ACKNOWLEDGMENTS}
We thank Marie Ericsson for discussions and useful comments.  
M.S. acknowledges financial support from the Swedish strategic research programme eSSENCE. 
E.S. thanks the Swedish Research Council (VR) for financial support through Grant No. 
D0413201. The computations were performed on resources provided by SNIC through Uppsala 
Multidisciplinary Center for Advanced Computational Science (UPPMAX) under Project 
snic2014-3-66.

\end{document}